\documentclass{article}
\usepackage{spconf,amsmath,graphicx}

\usepackage{enumitem}
\usepackage{multicol}
\usepackage[colorlinks=true,allcolors=blue]{hyperref}
\usepackage{url}
\usepackage{graphicx}
\usepackage{amsmath}
\usepackage{amssymb}
\usepackage{booktabs}
\usepackage{enumitem}
\usepackage{float}
\restylefloat{table}
\usepackage{stfloats}
\usepackage{graphicx}
\usepackage{bold-extra}
\usepackage{bm}
\usepackage{url}
\usepackage[utf8]{inputenc}
\usepackage{amsthm}
\usepackage{lipsum} 
\usepackage{booktabs}
\usepackage{algorithm}
\usepackage[utf8]{inputenc}
\usepackage{amssymb,color}
\usepackage{caption}
\usepackage{makecell,booktabs}
\usepackage{mathtools}
\usepackage{multicol}
\usepackage{amsmath, amssymb, dsfont}
\usepackage{mathabx}
\usepackage{epstopdf}
\usepackage{subcaption}
\usepackage{breqn}
\usepackage{verbatim}
\usepackage{arydshln}
\usepackage{tikz}
\usepackage{tikz,pgfplots}
\usepackage{pgfplots}
\usetikzlibrary{patterns}
\usepackage{algorithm}
\usepackage{algpseudocode}
\usepackage{amsfonts}

\title{Enabling Quartile-based Estimated-Mean Gradient Aggregation As Baseline for Federated Image Classifications}
\name{Yusen Wu, Jamie Deng, Hao Chen, Phuong Nguyen, Yelena Yesha}
\address{Dept. of Computer Science, University of Miami, Miami, Florida, USA}

\begin{document}
%
\maketitle
\begin{abstract}
Federated Learning (FL) has revolutionized how we train deep neural networks by enabling decentralized collaboration while safeguarding sensitive data and improving model performance. However, FL faces two crucial challenges: the diverse nature of data held by individual clients and the vulnerability of the FL system to security breaches. This paper introduces an innovative solution named Estimated Mean Aggregation (EMA) that not only addresses these challenges but also provides a fundamental reference point as a $\mathsf{baseline}$ for advanced aggregation techniques in FL systems. EMA's significance lies in its dual role: enhancing model security by effectively handling malicious outliers through trimmed means and uncovering data heterogeneity to ensure that trained models are adaptable across various client datasets. Through a wealth of experiments, EMA consistently demonstrates high accuracy and area under the curve (AUC) compared to alternative methods, establishing itself as a robust baseline for evaluating the effectiveness and security of FL aggregation methods. EMA's contributions thus offer a crucial step forward in advancing the efficiency, security, and versatility of decentralized deep learning in the context of FL. 
$\mathsf{\href{https://github.com/AnonymousUser08/EMA.git}{https://github.com/AnonymousUser08/EMA.git}}$
\end{abstract}

\section{Introduction}
Federated Learning (FL) \cite{kairouz2021advances} is a groundbreaking approach that revolutionizes model training by enabling collaborative learning across multiple devices or clients while maintaining localized data. Unlike traditional machine learning, where data is typically centralized for model training, FL allows training models directly on clients without transferring raw data to a central server. The aggregation of gradients computed by individual clients is used to update the global model. This methodology ensures data privacy while collectively enhancing model performance. However, the efficacy of FL faces challenges that require careful consideration.

A significant challenge is data heterogeneity \cite{chai2019towards}, which refers to variations in data distribution due to different sources or subsets having distinct distributions. Data can be either independent and identically distributed (IID) or non-IID \cite{zhao2018federated}. The intricate nature of data heterogeneity complicates gradient aggregation from different clients. Additionally, ensuring model security within federated learning systems is crucial due to the vulnerabilities of clients to various security breaches, such as poisoning attacks \cite{tian2022comprehensive}. Emerging threats like gradient inversion attacks \cite{huang2021evaluating} emphasize the need for robust defense mechanisms to prevent privacy breaches.

Numerous strategies, such as Median \cite{yin2018byzantine}, Trimmed mean \cite{yin2018byzantine}, Zeno \cite{xie2019zeno}, and Krum \cite{blanchard2017machine}, have been proposed to counteract attacks or Byzantine failures \cite{fang2020local}. However, these strategies often struggle to address data heterogeneity which occurs on the client side, while gradient aggregation occurs on the server side. Moreover, the lack of standardized baselines further complicates the evaluation of aggregation methods. 

To address these challenges, we propose an innovative estimated mean aggregation (EMA) methodology. EMA mitigates the impact of malicious outliers by utilizing trimmed mean and controlling the influence of outliers through the interquartile range (IQR) \cite{wan2014estimating}. This enhances model security by systematically handling outlier removal. Furthermore, our approach comprehensively evaluates trained models across non-IID or IID datasets, identifying data heterogeneity and assessing adaptability. This is achieved by assessing model performance on individual clients' datasets using mean squared error (MSE) \cite{koksoy2006multiresponse} to calculate the threshold for the coefficient of variation (CV) of the loss values \cite{abdi2010coefficient}. Our work contributes to the field in the following ways:
\begin{itemize}[itemsep=-1pt] 
\item We present a novel and efficient method for gradient aggregation, termed EMA. Our EMA estimated mean aggregation approach ensures the preservation of model security.
\item We developed a threshold-based CV algorithm to assess client-specific model adaptability and exclude certain client updates that may be from heterogeneous datasets.
\item We evaluated EMA in comparison to other advanced aggregation rules across varying biomedical datasets. The results show that EMA  outperforms its competitors in terms of model accuracy and AUC, as the baseline aggregation technique using estimated gradients.

\end{itemize}

\section{PROPOSED METHOD}
\label{sec:method}

In this section, we first provide the necessary definition of our quartile-based estimated mean. Next, we present all the modules that comprise EMA and discuss how we identify data heterogeneity from different clients. In order to explain our solution, we formalized our code in Algorithm \ref{alg:quartile}.

\subsection{Definition of Estimated Mean}
In this context, we emphasize the essential definition of the estimated sample mean as outlined in Definition 1. The formal proofs supporting this definition are available in the reference papers \cite{luo2018optimally}.

\noindent \textbf{Definition 1}. (Estimated Mean \cite{luo2018optimally}) For Scenario S = \{$q_1$, $m$, $q_3$; $N$\} where $q_1$ is the first quartile, $q_3$ is the third quartile, $m$ denotes the median, and $N$ stands for the sample size, we involve Luo's theorem and give the estimator $\bar X(w)$,
\begin{align}
\bar X(w) = w(q_1+q_3)/2 + (1-w)m
\label{equation1}
\end{align}
We have the following conclusions: \\ 
(i) $\bar X(w)$ is an unbiased estimator of $\mu$, i.e., $\mathbf{E}[\bar X(w)] =\mu$. \\
(ii) $\mathbf{E}[\bar X(w)-\mu] = \frac{w^2}{4}\textit{Var}[q_1+q_3] + (1-w)^2 \textit{Var}[m] +w(1-w)\textit{Cov}[m, q_1+q_3]$. \\ 
(iii) $w^* = \frac{\textit{Var}(m)-2\textit{Cov}(q_1+q_3,m)}{\textit{Var}(q_1+q_3)+4\textit{Var}(m)-4\textit{Cov}(q_1+q_3,m)}$. \\
(iv) When $N$ is large, $w^* \approx 0.699$. \\
Here, $Var$ and $Cov$ stand for variance and covariance. 

\begin{algorithm}[t!]
\footnotesize
\caption{Quartile-based Estimated-Mean Aggregation Rule}
\label{alg:quartile}
\begin{algorithmic}[1]
\Require Local updates $\{\Delta_{li}^T\}_{i=1}^T$, $\Delta_{l1}^{1},...,\Delta_{ln}^{T}$, lr, batch size (bs)
\Ensure  global update $\Delta_{g}^{t}$

\State \textit{/* Server */}
\State Authenticate client signatures
\State In epoch $t$, broadcast global update $\Delta_g^{t}$ to each client.
\State Wait for all $\Delta_{li}^{t+1}$ to arrive and collect them in array $\mathbb{C}$.
\State $\textbf{sg} \gets \text{sort}(\mathbb{C})$ \textit{/* sorted gradients */}
\If{$\text{length}(\textbf{sg}) \% 2 == 1$}
    \State $\textbf{m} \gets \textbf{sg}[\text{length}(\textbf{sg})/2]$ \textit{/* m refers to median */}
\Else
    \State $\textbf{m} \gets (\textbf{sg}[\text{length}(\textbf{sg})/2-1] + \textbf{sg}[\text{length}(\textbf{sg})/2])/2$
\EndIf
\State $\mathbf{G}_{\text{aggregated}} \gets \textbf{sg}[L:U]$ \textit{/* L:lower-bound, U: upper-bound*/}

\State \textit{/* Estimated mean */}
\State Epoch $t+1$:
\State  $\text{quar1} \gets \text{length}(\mathbf{G}_{\text{aggregated}})/4$
\State  $\text{quar3} \gets \text{quar1} \times 3$
\State  $w \gets 0.70+0.39/n$
\State  $\textbf{q1} \gets \mathbf{G}_{\text{aggregated}}[:\text{quar1}-1]$
\State  $\textbf{q3} \gets \mathbf{G}_{\text{aggregated}}[:\text{quar3}-1]$
\State   $\mathbf{G}_{\text{estimated}} \gets w \times (\textbf{q1} + \textbf{q3})/2 + (1-w) \times \textbf{m}$
\State  $\Delta^{t+1}_g  \gets \mathbf{G}_{\text{estimated}}$
\State  broadcast $\Delta^{t+1}_g$ to all clients as a global update.
\State \textit{/* Authorized Clients */}
\For{epoch $t = 1,\ldots, T$}
    \State Wait to receive $\Delta^{t}_g$ from the server.
    \State Train, compute, and send local update $\Delta^{t+1}_{li}$ to the server.
\EndFor
\end{algorithmic}
\end{algorithm}

\subsection{Quartile-based Estimated-Mean Gradient Aggregation Rule}
The Quartile-based Estimated-Mean Gradient Aggregation Rule stands as a cornerstone of this study, providing an innovative and efficient technique for aggregating gradients while preserving model security. EMA introduces a methodical process that facilitates the reliable estimation of mean gradients for decentralized training. To demonstrate the utilization of estimated aggregation (the gradients are in a normal distribution), we employ Shapiro-Wilk and Anderson-Darling \cite{razali2011power} tests to assess gradients across 50 clients. Furthermore, we delve into the capacity of EMA to accommodate outliers while upholding model security.

\noindent \textbf{Shapiro-Wilk and Anderson-Darling Test}
The Shapiro-Wilk and Anderson-Darling tests \cite{razali2011power} are statistical tests used to assess the normality of a dataset.  In order to run the statistical tests between 50 clients, we reshaped each of the client's tensors into a 2D tensor with one column and an unknown number of rows. A $\mathsf{torch.cat()}$ function is provided to concatenate (joining together) tensors along a specified dimension (50 columns). It combines multiple tensors into a single tensor. We then conducted 1,075,621 tests to validate our assumptions. The results indicate that in FL settings, over 78.3\% of the gradients follow a normal distribution across the CIFAR10 and MNIST benchmarks in one epoch (more medical datasets are employed in Section \ref{sec:eva}). In this paper, we named it the \textbf{Pre-Testing rate}. Importantly, different datasets may exhibit varying percentages of Pre-Testing rates due to the potential lack of IID data. Prior to training, it is necessary to conduct Pre-Testing to understand the proportion of gradients that may deviate from the normal distribution so that we can set up the borderline. We demonstrated that clients' gradients are in a normal distribution (e.g., more clients lead to a more accurate mean), meaning that the estimated mean algorithm is \underline{feasible} to estimate the global update. 

\noindent \textbf{Model Security and Outlier Handling}
One of EMA's primary advantages is its capacity to enhance model security by effectively handling malicious outliers during aggregation. By incorporating quartiles, EMA identifies and mitigates the influence of outliers that could compromise the aggregated gradient's integrity. This process is particularly valuable in guarding against adversarial manipulation and ensuring a trustworthy model update.

\begin{algorithm} [t!]
\footnotesize
\caption{Upper/Lower Threshold Settings}
\label{algorhtm2}
\begin{algorithmic}[1]
\Require Set of gradients: $\{ \mathbf{G}_1, \mathbf{G}_2, \ldots, \mathbf{G}_n \}$
\State Calculate first quartile ($q1$) and third quartile ($q3$) of gradient distribution
\State Calculate interquartile range (IQR): $I_{iqr} = q3 - q1$
\State Calculate lower outlier threshold: $\text{Lower Threshold} = q1 - k \times I_{iqr}$
\State Calculate upper outlier threshold: $\text{Upper Threshold} = q3 + k \times I_{iqr}$
\State Initialize aggregated gradient: $\mathbf{G}_{\text{aggregated}} = [0, 0, \ldots, 0]$ (matching dimensions)
\For{$i = 1$ to $n$}
    \If{$\mathbf{G}_i$ lies between Lower Threshold and Upper Threshold}
        \State Add $\mathbf{G}_i$ to $\mathbf{G}_{\text{aggregated}}$
    \EndIf
\EndFor
\State Normalize $\mathbf{G}_{\text{aggregated}}$ by dividing by the number of non-outlier gradients
\State \textbf{Output:} Normalized aggregated gradient: $\mathbf{G}_{\text{aggregated}}$
\end{algorithmic}
\end{algorithm}

We created an upper threshold and a lower threshold to normalize gradients by dividing the number of non-outlier gradients as shown in Algorithm \ref{algorhtm2}. We name it the normalized gradient $\mathbf{G}_{\text{aggregated}}$. We then run the quartile-based estimated-mean algorithm to get an estimated global update $\mathbf{G}_{\text{estimated}}$, and broadcast $\Delta^{t+1}_g$ to all clients as a global update, as shown in Algorithm \ref{alg:quartile}.

\subsection{Identifying Data Heterogeneity}

\begin{algorithm}[t!]
\footnotesize
\caption{Evaluate Trained Model on Clients}
\label{alg:evaluate_model_combined}
\begin{algorithmic}[1]
\Require List of clients $\{C_1, C_2, \ldots, C_n\}$, Trained model $\theta$
\Function{EvaluateModelOnClient}{$C_i$, $\theta$}
    \State $(D_i, T_i) \gets (C_i.\text{dataset}, C_i.\text{targets})$
    \State $\theta \gets \theta.\text{get()}$
    \State $\text{criterion} \gets \text{MSELoss()}$
    \State $O_i \gets \theta(D_i)$
    \State $\ell_i \gets \text{criterion}(O_i, T_i)$
    \State \Return $\ell_i.\text{item()}$
\EndFunction

\State

\For{each client $L_i$ in $\{C_1, C_2, \ldots, C_n\}$}
    \State $\ell_i \gets \text{EvaluateModelOnClient}(C_i, \theta)$
    \State \textbf{print} Client: $C_i.\text{id}$, Loss: $\ell_i$ rounded to four decimal places
\EndFor
\end{algorithmic}
\end{algorithm}

\begin{algorithm}[t!]
\footnotesize
\caption{Identifying Non-IID Datasets}
\label{alg:identify_non_iid}
\begin{algorithmic}[1]
\Require Loss values for clients $\{L_1, L_2, \ldots, L_n\}$
\Ensure Non-IID assessment

\State Calculate mean loss: $\mu \gets \frac{1}{n} \sum_{i=1}^{n} L_i$
\State Calculate standard deviation: $\sigma \gets \sqrt{\frac{1}{n} \sum_{i=1}^{n} (L_i - \mu)^2}$
\State Calculate coefficient of variation (CV): $\text{CV} \gets \frac{\sigma}{\mu}$

\State Set threshold: $d \gets \text{threshold}$

\If{$\text{CV} > d$}
    \State \textbf{Output} The dataset is likely non-iid.
\Else
    \State \textbf{Output} The dataset is likely iid.
\EndIf

\end{algorithmic}
\end{algorithm}

 The Algorithm \ref{alg:evaluate_model_combined} $\mathsf{EvaluateModelOnClient}$ is a subroutine used within the main Algorithm \ref{alg:quartile}. It takes specific inputs, performs calculations, and returns a value. The algorithm starts with a comment specifying the input requirements. In this case, it expects four inputs: (1), client $C_i$: Represents the client on which the model is being evaluated. (2), Model $\theta$: The trained model that will be used for evaluation. (3), Data $D_i$: The data to be input into the model for evaluation. (4), Targets $T_i$ (label): The target values corresponding to the data. It computes the loss between the model's predictions and the actual targets using the Mean Squared Error criterion and returns the calculated loss as a numeric value. 

By comparing the loss values from different clients, the server can gain insights into the distribution of data across clients. If the loss values vary significantly, it might indicate that the data from a certain client is non-IID. The server then can make decisions on whether to include or exclude certain client updates based on their loss values. In addition, We use the threshold of coefficient of variation (CV) to identify non-iid or iid datasets as shown in Algorithm \ref{alg:identify_non_iid}. If the loss between clients differs significantly, the CV value will be large.

\section{Evaluations}
\label{sec:eva}

\noindent \textbf{Data Preparation}.
Our evaluation of EMA encompasses a range of datasets, including the widely-used MNIST \cite{lecun1998gradient} and CIFAR-10 \cite{krizhevsky2009learning}. Our evaluation strategy extends to three distinct neural network architectures: ResNet18, ResNet34, and ResNet50. This comprehensive analysis showcases EMA's efficacy across a diverse set of model complexities. Expanding the scope of our investigation, we further evaluate the utility of our EMA approach on 6 2D biomedical datasets \cite{yang2023medmnist}: BloodMNIST, OrganAMNIST, OrganCMNIST, OrganSMNIST, PathMNIST, and DermaMNIST. This broader assessment emphasizes the applicability of EMA within various domains and datasets.

\noindent \textbf{Experimental Setup}.
We conducted the implementation and evaluation of EMA in a simulated environment using Tesla A100 Nvidia GPU settings. For our experiments, we configured the learning rate to 0.01 and utilized a batch size of 128. Additionally, our federated learning setup involved 50 clients participating in the training process.

\begin{figure}[ht!]
        \vspace{-6mm}
	\centering
        \begin{minipage}{\linewidth}
        \centering
        \begin{minipage}{1\linewidth}
          \begin{figure}[H]
              \includegraphics[width=\linewidth]{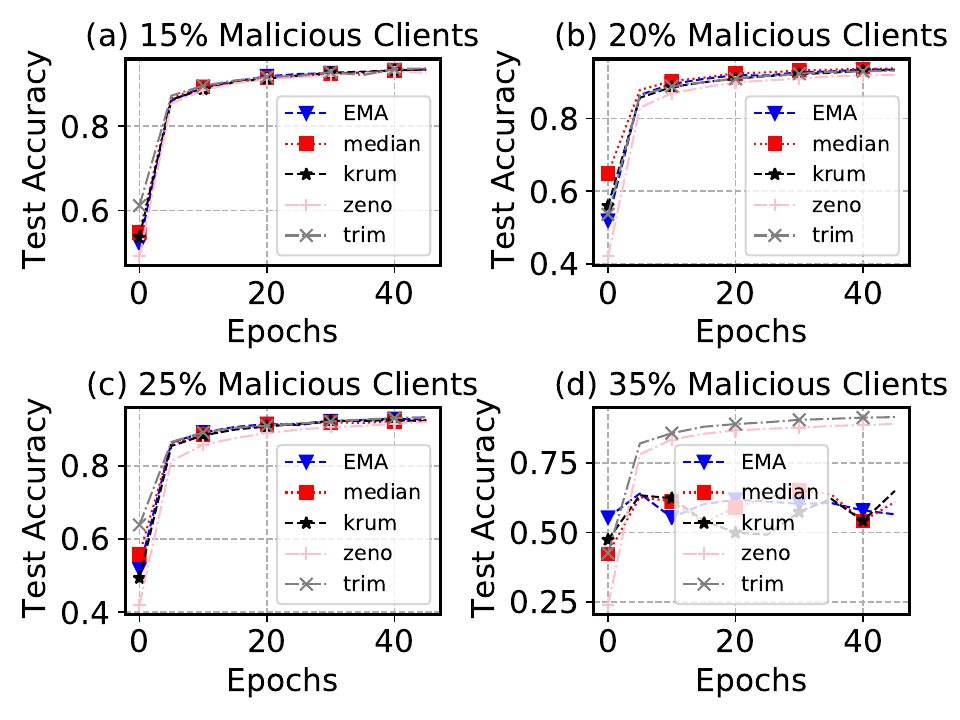}
              \caption{Different aggregation methods under CIFAR10 using a 6-layer neural network with varying percentages of attacks. }
              \label{fig:4}
          \end{figure}
        \end{minipage}
    
        \begin{minipage}{1\linewidth}
          \begin{figure}[H]
              \includegraphics[width=\linewidth]{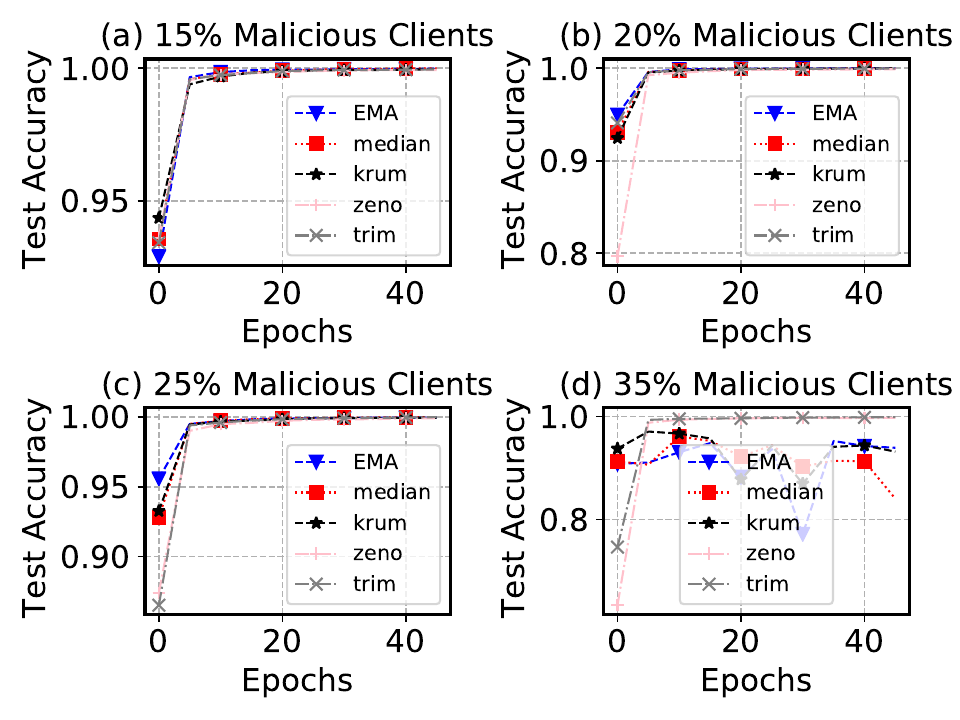}
              \caption{Different aggregation methods under MNIST using a 6-layer neural network with varying percentages of attacks.}
              \label{fig:5}
          \end{figure}
        \end{minipage}
  \end{minipage}
\end{figure}

\noindent \textbf{Assessing EMA Accuracy Across Varied Attack Percentages}. We utilize the estimated mean as the global updates to train the model and subsequently assess the test accuracy in the presence of various attacks. This evaluation is performed to compare the performance of our approach with other aggregation rules. We demonstrate that using the estimated mean as the global update also yields reliable performance, as shown in Fig. \ref{fig:4} and Fig. \ref{fig:5}.

Our tests revealed that when the percentage of harmful attacks is less than 25\%, the accuracy of our estimated mean is similar to the trimmed mean and Zeno on the CIFAR-10 dataset. Similarly, on the MNIST dataset, our estimated mean works best when malicious attacks are below 25\%. As the percentage of malicious clients increases, the accuracy of our estimated mean decreases, but it can still be used as a reference point to identify malicious clients. For instance, when the accuracy drops to around 60\%, it suggests that 
 there are around 35\% of malicious clients among all clients.

\noindent \textbf{Evaluating EMA Across Various Models}.
We conducted an evaluation of our EMA technique using three distinct models and 6 biomedical datasets, as detailed in Table \ref{tab:model_evaluation}. In comparison to the official evaluation results presented in \cite{yang2023medmnist}, our EMA approach demonstrated competitive performance. Notably, for certain cases, our method achieved test accuracies that exceeded the benchmark results, as indicated by the highlighted entries in the table.   

\begin{table}[h]
    \centering
    \footnotesize
    \begin{tabular}{@{} lcc lcc lcc lcc lcc lcc @{}}
        \toprule
        Datasets & \multicolumn{2}{c}{ResNet18} & \multicolumn{2}{c}{ResNet34} & \multicolumn{2}{c}{ResNet50} \\
        \cmidrule(r){2-3} \cmidrule(lr){4-5} \cmidrule(lr){6-7} \cmidrule(lr){8-9} \cmidrule(l){10-11}
        & ACC & AUC & ACC & AUC & ACC & AUC \\
        \midrule
        Path & 0.88 & 0.98 & 0.79 & 0.96 & 0.83 & 0.97 \\
        Derma & 0.70 & 0.87 & \textbf{0.75} & 0.89 & 0.73 & 0.90 \\
        Blood & 0.87 & 0.99 & 0.83 & 0.98 & 0.86 & 0.98 \\
        OrganA & 0.84 & 0.99 & 0.83 & 0.98 & 0.87 & 0.99 \\
        OrganC & 0.81 & 0.98 & 0.83 & 0.98 & \colorbox{gray!20}{\textbf{0.84}} & 0.98 \\
        OrganS & 0.69 & 0.96 & 0.67 & 0.95 & \colorbox{gray!20}{\textbf{0.71}} & 0.96 \\
        \bottomrule
    \end{tabular}
    \caption{Evaluation results of EMA for accuracy (ACC) and area under the curve (AUC) with 6 different medical datasets.}
    \label{tab:model_evaluation}
\end{table}

\noindent \textbf{Comparative Evaluation of EMA with Various Aggregation Rules}. We conducted an evaluation of our EMA approach and compared its performance with four other popular aggregation methods: Median, Zeno, Krum, and Trimmed Mean under the OrganCMINIST dataset. This evaluation was conducted without the inclusion of malicious attacks. The results of our study indicate that EMA demonstrates a similar level of performance, with certain accuracy metrics even better than those of the commonly used aggregation methods.

We claim that EMA can be a valuable baseline in FL settings. The EMA can provide reasonable initial results without requiring extensive parameter tuning or complex adjustments (e.g., only 3 points to estimate the mean). This makes it a good starting point for building more advanced aggregation techniques.

\begin{table}[h]
    \centering
    \footnotesize
    \begin{tabular}{@{} lcc lcc lcc lcc lcc lcc @{}}
        \toprule
        Aggr & \multicolumn{2}{c}{ResNet18} & \multicolumn{2}{c}{ResNet34} & \multicolumn{2}{c}{ResNet50} \\
        \cmidrule(r){2-3} \cmidrule(lr){4-5} \cmidrule(lr){6-7} \cmidrule(lr){8-9} \cmidrule(l){10-11}
        & ACC & AUC & ACC & AUC & ACC & AUC \\
        \midrule
        Median & 0.83 & 0.98 & 0.82 & 0.98 & 0.83 & 0.98 \\
        Zeno & 0.80 & 0.98 & 0.80 & 0.98 & 0.80 & 0.98 \\
        Krum & 0.81 & 0.98 & 0.81 & 0.97 & 0.82 & 0.98 \\
        Trim & \colorbox{gray!20}{\textbf{0.83}} & 0.98 & 0.83 & 0.98 & 0.81 & 0.98 \\
        \textbf{EMA} & 0.81 & 0.98 & \textbf{0.83} & 0.98 & \colorbox{gray!20}{\textbf{0.84}} & 0.98 \\
        \bottomrule
    \end{tabular}
    \caption{Model evaluation results for test accuracy (ACC) and area under the curve (AUC) with 4 different aggregation methods under the OrganCMINIST dataset.}
    \label{tab:model_evaluation2}
\end{table}
\noindent \textbf{Evaluating Data Heterogeneity} To identify non-iid datasets, we use statistical measures (Algorithm \ref{alg:identify_non_iid}) that highlight variations in loss values across clients. Our approach is to calculate the coefficient of variation (CV) or standard deviation of the loss values. High CV or standard deviation values indicate non-iid data since different datasets can have varying characteristics that impact the loss. 

We evaluated 8 medical datasets where we found that if one dataset has a higher proportion of biased data or outliers, the loss performs worse. Empirically, a smaller threshold value $d$ implies more restriction in identifying non-iid.  For instance, considering the optimal loss values for Non-IID MNIST \cite{zhao2018federated} and IID MNIST as 2.4 and 1.4 respectively, a threshold of 0.25 can be strategically set to effectively facilitate identification.

\section{Conclusions}
We propose EMA as a robust aggregation solution for FL. We also propose to use a threshold of the coefficient of variation to identify data heterogeneity. FL's privacy-focused collaborative approach, coupled with EMA and our algorithm, presents a promising baseline for efficient and secure model training across different clients in FL.
\bibliographystyle{IEEEbib}
\bibliography{strings,refs}

\begin{thebibliography}{10}

\bibitem{kairouz2021advances}
Peter Kairouz, H~Brendan McMahan, Brendan Avent, Aur{\'e}lien Bellet, Mehdi
  Bennis, Arjun~Nitin Bhagoji, Kallista Bonawitz, Zachary Charles, Graham
  Cormode, Rachel Cummings, et~al.,
\newblock ``Advances and open problems in federated learning,''
\newblock {\em Foundations and Trends{\textregistered} in Machine Learning},
  vol. 14, no. 1--2, pp. 1--210, 2021.

\bibitem{chai2019towards}
Zheng Chai, Hannan Fayyaz, Zeshan Fayyaz, Ali Anwar, Yi~Zhou, Nathalie
  Baracaldo, Heiko Ludwig, and Yue Cheng,
\newblock ``Towards taming the resource and data heterogeneity in federated
  learning,''
\newblock in {\em 2019 USENIX conference on operational machine learning (OpML
  19)}, 2019, pp. 19--21.

\bibitem{zhao2018federated}
Yue Zhao, Meng Li, Liangzhen Lai, Naveen Suda, Damon Civin, and Vikas Chandra,
\newblock ``Federated learning with non-iid data,''
\newblock {\em arXiv preprint arXiv:1806.00582}, 2018.

\bibitem{tian2022comprehensive}
Zhiyi Tian, Lei Cui, Jie Liang, and Shui Yu,
\newblock ``A comprehensive survey on poisoning attacks and countermeasures in
  machine learning,''
\newblock {\em ACM Computing Surveys}, vol. 55, no. 8, pp. 1--35, 2022.

\bibitem{huang2021evaluating}
Yangsibo Huang, Samyak Gupta, Zhao Song, Kai Li, and Sanjeev Arora,
\newblock ``Evaluating gradient inversion attacks and defenses in federated
  learning,''
\newblock {\em Advances in Neural Information Processing Systems}, vol. 34, pp.
  7232--7241, 2021.

\bibitem{yin2018byzantine}
Dong Yin, Yudong Chen, Ramchandran Kannan, and Peter Bartlett,
\newblock ``Byzantine-robust distributed learning: Towards optimal statistical
  rates,''
\newblock in {\em Proc. International Conference on Machine Learning}. PMLR,
  2018, pp. 5650--5659.

\bibitem{xie2019zeno}
Cong Xie, Sanmi Koyejo, and Indranil Gupta,
\newblock ``Zeno: Distributed stochastic gradient descent with suspicion-based
  fault-tolerance,''
\newblock in {\em International Conference on Machine Learning}. PMLR, 2019,
  pp. 6893--6901.

\bibitem{blanchard2017machine}
Peva Blanchard, El~Mahdi El~Mhamdi, Rachid Guerraoui, and Julien Stainer,
\newblock ``Machine learning with adversaries: Byzantine tolerant gradient
  descent,''
\newblock {\em Advances in Neural Information Processing Systems}, vol. 30,
  2017.

\bibitem{fang2020local}
Minghong Fang, Xiaoyu Cao, Jinyuan Jia, and Neil Gong,
\newblock ``Local model poisoning attacks to $\{$Byzantine-Robust$\}$ federated
  learning,''
\newblock in {\em Proc. USENIX Security Symposium (USENIX Security 20)}, 2020,
  pp. 1605--1622.

\bibitem{wan2014estimating}
Xiang Wan, Wenqian Wang, Jiming Liu, and Tiejun Tong,
\newblock ``Estimating the sample mean and standard deviation from the sample
  size, median, range and/or interquartile range,''
\newblock {\em BMC medical research methodology}, vol. 14, no. 1, pp. 1--13,
  2014.

\bibitem{koksoy2006multiresponse}
Onur K{\"o}ksoy,
\newblock ``Multiresponse robust design: Mean square error (mse) criterion,''
\newblock {\em Applied Mathematics and Computation}, vol. 175, no. 2, pp.
  1716--1729, 2006.

\bibitem{abdi2010coefficient}
Herv{\'e} Abdi,
\newblock ``Coefficient of variation,''
\newblock {\em Encyclopedia of research design}, vol. 1, no. 5, 2010.

\bibitem{luo2018optimally}
Dehui Luo, Xiang Wan, Jiming Liu, and Tiejun Tong,
\newblock ``Optimally estimating the sample mean from the sample size, median,
  mid-range, and/or mid-quartile range,''
\newblock {\em Statistical methods in medical research}, vol. 27, no. 6, pp.
  1785--1805, 2018.

\bibitem{razali2011power}
Nornadiah~Mohd Razali, Yap~Bee Wah, et~al.,
\newblock ``Power comparisons of shapiro-wilk, kolmogorov-smirnov, lilliefors
  and anderson-darling tests,''
\newblock {\em Journal of statistical modeling and analytics}, vol. 2, no. 1,
  pp. 21--33, 2011.

\bibitem{lecun1998gradient}
Yann LeCun, L{\'e}on Bottou, Yoshua Bengio, and Patrick Haffner,
\newblock ``Gradient-based learning applied to document recognition,''
\newblock {\em Proceedings of the IEEE}, vol. 86, no. 11, pp. 2278--2324, 1998.

\bibitem{krizhevsky2009learning}
Alex Krizhevsky, Geoffrey Hinton, et~al.,
\newblock ``Learning multiple layers of features from tiny images,''
\newblock 2009.

\bibitem{yang2023medmnist}
Jiancheng Yang, Rui Shi, Donglai Wei, Zequan Liu, Lin Zhao, Bilian Ke,
  Hanspeter Pfister, and Bingbing Ni,
\newblock ``Medmnist v2-a large-scale lightweight benchmark for 2d and 3d
  biomedical image classification,''
\newblock {\em Scientific Data}, vol. 10, no. 1, pp. 41, 2023.

\end{thebibliography}

\end{document}